\documentclass[aps,reprint,twocolumn,superscriptaddress,citeautoscript,amsmath,amssymb,floatfix,longbibliography,prb,]{revtex4-2}
\usepackage{graphicx}
\usepackage{times}
\usepackage{bm}
\usepackage{color}
\usepackage{gensymb}
\usepackage{dcolumn}
\usepackage{xfrac}

\newcolumntype{d}[1]{D{.}{.}{#1}}

\begin{document}
	\title{Europium $c$-axis ferromagnetism in Eu(Co$_{\bm{1-x}}$Ni$_{\bm{x}}$)$_{\bm{2-y}}$As$_{\bm{2}}$: A single-crystal neutron diffraction study}
	
	\author{Tianxiong~Han}
	\affiliation{Division of Materials Sciences and Engineering, Ames National Laboratory, Ames, Iowa 50011, USA}
	\affiliation{Department of Physics and Astronomy, Iowa State University, Ames, Iowa 50011, USA}
	
	\author{Santanu~Pakhira}
	\altaffiliation{Institute for Quantum Materials and Technologies, Karlsruhe Institute of Technology, 76131, Karlsruhe, DE}
	\affiliation{Division of Materials Sciences and Engineering, Ames National Laboratory, Ames, Iowa 50011, USA}
	
	\author{N.~S.~Sangeetha}
	\affiliation{Division of Materials Sciences and Engineering, Ames National Laboratory, Ames, Iowa 50011, USA}
	
	\author{S.~X.~M.~Riberolles}
	\affiliation{Division of Materials Sciences and Engineering, Ames National Laboratory, Ames, Iowa 50011, USA}
	
	\author{T.~W.~Heitmann}
	\affiliation{University of Missouri Research Reactor,  University of Missouri, Columbia, MO, 65211, USA}
	\affiliation{Department of Physics and
		Astronomy, University of Missouri, Columbia, MO, 65211, USA}
	\affiliation{MU Materials Science and Engineering Institute, University of Missouri, Columbia, MO, 65211, USA}
	
	\author{Yan~Wu}
	\affiliation{Neutron Scattering Division, Oak Ridge National Laboratory, Oak Ridge, Tennessee, 37831, USA}
	
	\author{D.~C.~Johnston}
	\affiliation{Division of Materials Sciences and Engineering, Ames National Laboratory, Ames, Iowa 50011, USA}
	\affiliation{Department of Physics and Astronomy, Iowa State University, Ames, Iowa 50011, USA}
	
	\author{R.~J.~McQueeney}
	\affiliation{Division of Materials Sciences and Engineering, Ames National Laboratory, Ames, Iowa 50011, USA}
	\affiliation{Department of Physics and Astronomy, Iowa State University, Ames, Iowa 50011, USA}
	
	\author{B.~G.~Ueland}
	\email{bgueland@ameslab.gov}
	\affiliation{Division of Materials Sciences and Engineering, Ames National Laboratory, Ames, Iowa 50011, USA}
	\affiliation{Department of Physics and Astronomy, Iowa State University, Ames, Iowa 50011, USA}
	
	\date{\today}
	
	\begin{abstract}
		We report neutron diffraction results for the body-centered-tetragonal series Eu(Co$_{1-x}$Ni$_x$)$_{2-y}$As$_2$, $x=0.10$, $0.20$, $0.42$, and $0.82$, $y\leq0.10$, that detail changes to the magnetic ordering with nominal hole doping. We report the antiferromagnetic (AFM) propagation vectors, magnetic transition temperatures, and the ordered magnetic moments.  We find a nonmonotonic change of the AFM propagation vector with $x$, with a minimum occurring at the tetragonal to collapsed-tetragonal phase crossover. For $x=0.10$ and $0.82$ we find $c$-axis helix ordering of the Eu magnetic moments (spins) similar to $x=0$ and $1$, with the spins oriented within the $\mathbf{ab}$-plane. For $x=0.20$ and $0.42$ we find  higher-temperature $c$-axis FM order and lower-temperature $c$-axis cone order. Using the extinction conditions for the space group, we discovered that the Eu spins are ordered in the higher-temperature $c$-axis FM phase for intermediate values of $x$, contrary to a previous report suggesting only Co/Ni spin ordering. Although we cannot directly confirm that the Co/Ni spins are also ordered, we suggest that $c$-axis itinerant-FM ordering of the Co/Ni spins could provide a molecular field that drives FM ordering of the Eu spins, which in turn provides the anisotropy for the lower-temperature $c$-axis cone order.
	\end{abstract}
	
	\maketitle
	
	\section{Introduction}
	
	Magnetic transition-metal pnictides have been a versatile and enlightening set of compounds, exhibiting, for example, remarkable local-moment and itinerant magnetism, robust magnetostructural coupling, and unconventional phases such as high-$T_{\text{c}}$ superconductivity \cite{Johnston_2010, Canfield_2010, Stewart_2011,Johnston_2011, Dai_2015, Ueland_2015, LiY_2019b,Fernandes_2022}.  Studies on $AM_2X_2$ ($122$) compounds, where $A$ is an alkaline-earth or rare-earth metal, $M$ is a transition metal, and $X$ is a pnictogen, have born out particularly intriguing phenomena. The tetragonal $A=$~Ca, Sr, Ba, $M=$~Fe, $X=$~As compounds exhibit weak-moment ($\lesssim 1~\mu_{\text{B}}~\text{per Fe}$) stripe-type antiferromagnetic (AFM) ordering of the Fe magnetic moments (spins) with the ordered moments lying in the $\mathbf{ab}$ plane, a tetragonal-to-orthorhombic structural transition, and carrier-doping-induced superconductivity with significant stripe-type AFM spin fluctuations.  For tetragonal BaMn$_2$As$_2$, local-moment N\'eel-type AFM ordering of the Mn spins is found with the ordered moment lying along the $\mathbf{c}$ axis. But for the hole-doped series Ba$_{1-x}$K$_x$Mn$_2$As$_2$, sufficient hole doping causes itinerant ferromagnetic (FM) order of spins in the As $4p$ bands with an ordered moment lying perpendicular to $\mathbf{c}$ which coexists with the local-moment AFM order of the Mn spins \cite{Pandey_2013, Ueland_2015}.
	
	The tetragonal $A$Co$_2$As$_2$, $A=$~Ca, Sr, Ba, compounds have different electronic ground states than the corresponding Fe $122$ compounds but also exhibit weak ordered magnetic moments and other properties suggestive of itinerant magnetism. CaCo$_{1.86}$As$_2$ displays A-type AFM ordering below a N\'eel temperature $T_{\text{N}}=51$~K with an ordered magnetic moment  $\mu=0.43(5)~\mu_{\text{B}}/\text{Co}$ lying along $\mathbf{c}$ \cite{Quirinale_2013, Jayasekara_2017}. On the other hand, magnetic ordering has yet to be found for SrCo$_2$As$_2$ (above $T=0.05$~K), which shows spin fluctuations similar to those seen in the  stripe-type AFM phases of the Fe $122$ compounds \cite{Jayasekara_2013,Li_2019}. In fact, the magnetic excitation spectra for the Ca and Sr compounds are both remarkable despite the compounds' different magnetic ground states; both spectra reflect the presence of frustration, magnetic phase competition, and itinerant magnetism  \cite{Sapkota_2017,LiY_2019,LiY_2019b,Li_2019,Ueland_2021}. For BaCo$_2$As$_2$, the compound has been reported to be a strongly-renormalized paramagnet lying close to a quantum critical point \cite{Sefat_2009,Pakhira_2021}, but obtaining large enough single-crystal samples for inelastic neutron scattering experiments has proven challenging \cite{Anand_2014a}.
	
	Replacing $A$ by a magnetic rare earth offers a seemingly straightforward way to insert $4f$ localized spins in the presence of itinerant Co spins and provide new pathways to tune the magnetic properties \cite{Marchand_1978, Zapf_2017}. However, it is not guaranteed that Co itinerant magnetism exists in the compound. For example, body-centered-tetragonal (Tet) EuCo$_{2-y}$As$_2$ exhibits $c$-axis helix AFM order of its localized Eu spins with no detectable magnetic ordering of its Co spins \cite{Ding_2017,Tan_2016,Sangeetha_2018}. EuNi$_{2-y}$As$_2$ exists in the collapsed-tetragonal (cT) phase at ambient pressure with an $\approx11$\% smaller $c$ lattice parameter than EuCo$_{2-y}$As$_2$ \cite{Sangeetha_2019, Sangeetha_2020}, and it also exhibits $c$-axis helix AFM order of the Eu spins as well as no ordering of the $3d$ Ni spins \cite{Jin_2019}.  A previous study of the thermodynamic and transport properties of Eu(Co$_{1-x}$Ni$_x$)$_{2-y}$As$_2$, however, has reported the emergence of itinerant ferromagnetic (FM) order of the Co/Ni spins for intermediate values of $x$ \cite{Sangeetha_2020}.  Here, we present neutron diffraction results for Eu(Co$_{1-x}$Ni$_x$)$_{2-y}$As$_2$ that directly determine changes to the magnetic ordering with $x$.
	
\begin{figure}[]
		\centering
		\includegraphics[width=1.0\linewidth]{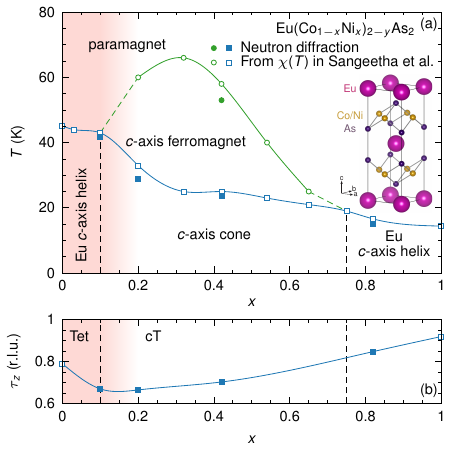}
		\caption{  \label{Fig:Phase_Dia} (a) Temperature versus composition magnetic phase diagram for Eu(Co$_{1-x}$Ni$_x$)$_2$As$_2$ from our neutron diffraction data and previously reported magnetic susceptibility data \cite{Sangeetha_2020}. The ThCr$_2$Si$_2$-type body-centered-tetragonal chemical-unit cell with space group $I4/mmm$ and room-temperature lattice parameters of  $a=3.9478(7)$~\AA\ and $c=11.232(2)$~\AA\ for $x=0$ \cite{Sangeetha_2018}, is also shown. (b) The component of the antiferromagnetic propagation vector $\bm{\tau}=(0,0,\tau_z)$ versus composition. The $x=0$ point is from Ref.~[\onlinecite{Tan_2016}] and the $x=1$ point is from Ref.~[\onlinecite{Jin_2019}]. Dashed lines are approximate phase boundaries and solid lines are guides for the eye. The shading illustrates the crossover between the tetragonal (Tet) and collapsed-tetragonal (cT) structures occurring between $x=0.10$ and $0.20$ \cite{Sangeetha_2020}.  r.l.u.\ stands for reciprocal-lattice unit.} 
	\end{figure}

	The chemical unit cell for Eu(Co$_{1-x}$Ni$_x$)$_{2-y}$As$_2$, where $y$ indicates vacancies of $y\leq0.10$  \cite{Sangeetha_2020}, is shown in Fig.~\ref{Fig:Phase_Dia}(a). The substitution of Ni for Co is expected to dope electrons into EuCo$_{2-y}$As$_2$ and provide a tuning knob for the electronic band structure and magnetic order. For example, the related compound tetragonal EuFe$_2$As$_2$ exhibits higher-temperature stripe-type ordering of its itinerant Fe spins and lower-temperature A-type ordering of its Eu spins with weak coupling between the two magnetic sublattices \cite{Xiao_2009, Martin_2009}. However, nominal hole doping by substitution of Co for Fe results in the suppression of the stripe-type itinerant AFM order and the emergence of superconductivity in Eu(Fe$_{0.89}$Co$_{0.11}$)$_2$As$_2$ \cite{Jiang_2009}. The superconductivity is destroyed upon cooling by the helical ordering of the Eu spins. Suppression of the Fe ordering in EuFe$_2$As$_2$ by pressure \cite{Miclea_2009} or substitution of P for As \cite{Jeevan_2011} has also been found to induce superconductivity. For the applied pressure study, superconductivity was reported to disappear once the Eu spins order. However, for the P-doped study, a region of coexisting Eu-spin AFM ordering and superconductivity was found. 
	
	For EuCo$_{2-y}$As$_2$, applying $p\approx4.7$~GPa of hydrostatic pressure causes a structural transition to the cT phase of the ThCr$_2$Si$_2$-type structure, which is generally characterized by an $\approx9$--$11\%$ decrease in $c$ and stronger As-As covalent bonding along $\mathbf{c}$ \cite{Hoffmann_1985}. The pressure-induced Tet-cT transition in EuCo$_{2-y}$As$_2$ is reported to be continuous (second order) and, for the cT phase, pressure increases the Eu valence to $2.25$.  This, in turn, induces itinerant-FM ordering of the Co spins due to the resulting electron doping of the $3d$ density of states \cite{Tan_2016}. Since substituting Ni for Co is another approach to electron dope EuCo$_{2-y}$As$_2$, it is interesting to examine the magnetic ordering across the Eu(Co$_{1-x}$Ni$_x$)$_{2-y}$As$_2$ series and examine whether a Tet-cT phase transition occurs as a function of $x$. Eu$^{2+}$ also is an interesting choice for $A$, as it has a large angular momentum of $J=7/2$ with $L=0$ orbital-angular momentum.  This latter fact means that there is no crystalline-electric-field (CEF) splitting of the single-ion ground-state angular-momentum multiplet which would establish magnetic anisotropy of the Eu spins. 
	
	The magnetic phase transitions across the Eu(Co$_{1-x}$Ni$_x$)$_{2-y}$As$_{2}$ series have been previously examined using magnetization, magnetic susceptibility, heat capacity, resistivity, x-ray diffraction, and  $^{151}$Eu M\"ossbauer spectroscopy \cite{Sangeetha_2020}. Results from these experiments determined a magnetic phase diagram where details of the magnetic order were largely inferred from the thermodynamic and transport data and complementary mean-field-theory calculations. The magnetic phase diagram is similar to the one in Fig.~\ref{Fig:Phase_Dia}(a), with two bookending regions of $c$-axis helix order of the Eu spins with no ordering of the Co/Ni spins. For the helix, each Eu layer contains ferromagnetically aligned spins laying in the $\mathbf{ab}$ plane and the orientation of spins in neighboring layers along $\mathbf{c}$ are rotated by the helix turn angle.  The middle of the phase diagram was inferred to feature a higher-temperature phase with itinerant $c$-axis FM ordering of the Co/Ni spins and a lower-temperature AFM phase in which the Eu spins adopt a $c$-axis cone structure.  The cone is similar to the $c$-axis helix, but the ordered Eu spins also have a component pointing along $\mathbf{c}$. A $2\tau$-helix region, which is not shown in Fig.~\ref{Fig:Phase_Dia}(a), was also inferred to exist for $0.02\alt x \alt 0.1$ with the AFM order having two AFM propagation vectors, one along $\mathbf{c}$ and one lying within the $\mathbf{ab}$ plane \cite{Sangeetha_2020}. Our neutron diffraction experiments did not find evidence for this phase and we omit it from Fig.~\ref{Fig:Phase_Dia}(a).
	
	Below, we report results from neutron-diffraction measurements on single-crystal samples of Eu(Co$_{1-x}$Ni$_x$)$_{2-y}$As$_2$ with $x=0.10$, $0.20$, $0.42$, and $0.82$.  In addition to finding an AFM propagation vector that changes nonmonotonically with $x$, we performed single-crystal refinements using the diffraction data  and directly determined the structure of the magnetic order in each phase, including the values of the ordered magnetic moments. We found low-temperature $c$-axis helix order for $x=0.10$ and $0.82$ and $c$-axis cone order for $x=0.20$ and $0.42$. We verified the existence of a higher-temperature $c$-axis FM phase for compositions exhibiting the $c$-axis cone at low temperature. However, we discovered that the Eu spins are ordered in the FM phase and that if the Co/Ni spins are also ordered, their ordered magnetic moment is $\mu_{\text{Co/Ni}}\lesssim0.2~\mu_{\text{B}}$. We discuss the neutron-diffraction results in the context of the previous thermodynamic, resistance, and M\"ossbauer spectroscopy results, and examine reasons for the ordered Eu spins to develop a component along $\mathbf{c}$ for intermediate values of $x$.
	
	\section{Experimental Details}
	Plate-like single crystals of Eu(Co$_{1-x}$Ni$_x$)$_{2-y}$As$_2$ with $x=0.10$ ($y=0.08$), $0.20$ ($y=0.06$), $0.42$ ($y=0.06$), and $0.82$ ($y=0.06$) were synthesized by solution growth and their compositions were determined as described previously \cite{Sangeetha_2020}.  Measurements of the magnetization $M$ were performed on each crystal while applying a magnetic field of $H=100$~Oe using a Quantum Design, Inc., superconducting quantum-interference device in order to compare the magnetic transition temperatures with those previously reported \cite{Sangeetha_2020}.  Data were taken down to a temperature of $T=2$~K, and the orientations of the single crystals were determined by x-ray Laue diffraction.
	
	Neutron diffraction experiments were performed on $x=0.10$, $0.42$, and $0.82$ single-crystal samples at the High-Flux Isotope Reactor at Oak Ridge National Laboratory using the $4$-circle neutron diffractometer on the DEMAND beamline.  Each sample was aligned such that the $(h,h,l)$ reciprocal-lattice plane was initially laid horizontally. This allowed us to achieve the best resolution for the $(h,h,l)$ plane while utilizing the $4$-circle mode of DEMAND. The $(2,2,0)$ reflection from a multilayer-$[1,1,0]$ wafer silicon monochromator was used to select a neutron wavelength of $\lambda=1.542$~\AA\ and a pyrolitic graphite (PG) filter was inserted into the beam to suppress higher-order wavelengths. A vacuum-chamber bottom-loading He closed-cycle-refrigerator (CCR) for the $4$-circle diffractometer was used for low-temperature control.
	
	A neutron diffraction experiment on an $x=0.20$ single-crystal sample was performed at the University of Missouri Research Reactor using the TRIAX triple-axis neutron spectrometer.  The sample was aligned with the $(h,h,l)$ reciprocal-lattice plane coincident with the horizontal scattering plane. The instrument was operated in elastic mode, using the $(0,0,2)$ Bragg reflection of a PG (PG$002$) monochromator and a PG$002$ analyzer.  A neutron energy of $E=30.5$~meV ($\lambda=1.638$~\AA) was selected to partially mitigate the significant thermal neutron absorption by Eu, and PG filters were inserted before and after the sample to suppress higher-order wavelengths.  S\"{o}ller slit collimators with divergences of $60^{\prime}$-$60^{\prime}$-$80^{\prime}$-$80^{\prime}$ were placed before the monochromator, between the monochromator and sample, between the sample and analyzer, and between the analyzer and detector, respectively. A bottom-loading CCR provided temperature control.
	
	Each neutron-diffraction sample was mounted to an Al sample holder using Al foil, Al wire, or glue, and the sample holder was subsequently mounted in an Al can.  The can was then evacuated, back-filled with He exchange gas, and sealed. Each can was thermally anchored to the cold head of the CCR used. Nuclear and magnetic structure-factor calculations and single-crystal refinements were made using \textsc{fullprof} \cite{fullprof} and corrections to the data to account for the thermal-neutron absorption of the samples were made using \textsc{mag2pol} \cite{Mag_2_pol}. Some diagrams were made using \textsc{vesta} \cite{Momma_2011}.
	
	\section{Results}
	\begin{figure}[]
		\centering
		\includegraphics[width=1.0\linewidth]{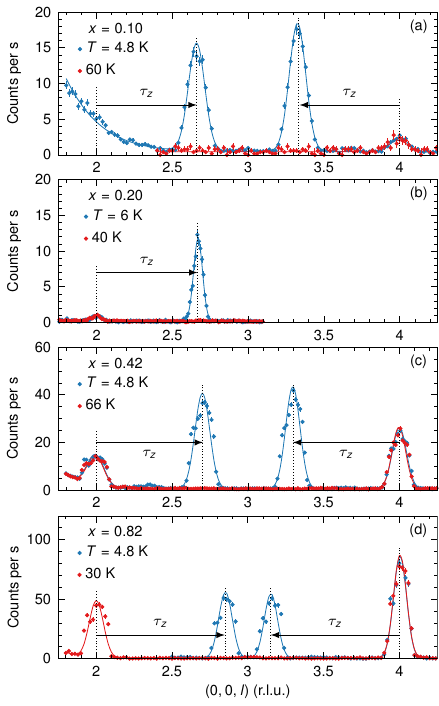}
		\caption{  \label{Fig:l_scans} $(0,0,l)$ neutron diffraction patterns for a temperature $T$ above and below the N\'{e}el temperature for $x=0.10$ (a), $0.20$ (b), $0.42$ (c), and $0.82$ (d). Lines show fits to Gaussian lineshapes and an empirical background. Note that the $y$-axis scale is different for each panel and that the background for $x=0.20$ [panel (b)] is different from the other panels. This is because the $x=0.20$ sample was measured on the TRIAX instrument whereas the other samples were measured on DEMAND.} 
	\end{figure}
	
	Each sample was cooled down to the base temperature of the CCR used and searches for magnetic-Bragg peaks were made using scans along $(0,0,l)$. Figures~\ref{Fig:l_scans}(a) to \ref{Fig:l_scans}(d) show $(0,0,l)$ neutron-diffraction patterns taken at the base temperatures and above $T_{\text{N}}$ for $x=0.10$, $0.20$, $0.42$, and $0.82$. The appearance of magnetic-Bragg peaks centered at noninteger values of $l$ is obvious in the lower-temperature data and is consistent with the presence of incommensurate AFM order. Structural (nuclear) Bragg peaks occur at $(0,0,l)$, $l$ even, positions and the distance from a structural-Bragg peak to the closest magnetic-Bragg peak determines the AFM propagation vector $\bm{\tau}=(0,0,\tau_{z})$.  A plot of $\tau_{z}(x)$ is shown in Fig.~\ref{Fig:Phase_Dia}(b), where a minimum is observed around $0.1\lesssim x\lesssim0.2$. For $x=0.1$, we did not find magnetic-Bragg peaks corresponding to a second AFM propagation vector despite performing a targeted search. Finally, measurements of multiple $(h,h,l)$ structural-Bragg peaks for each sample, which were used for the single-crystal refinements discussed below, gave results consistent with the previously-reported $I4/mmm$ space group and lattice parameters for the chemical lattice \cite{Sangeetha_2018}. 
	
	\begin{figure}[]
		\centering
		\includegraphics[width=1.0\linewidth]{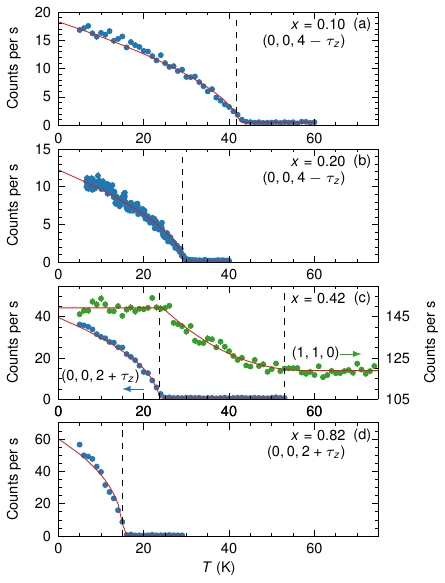}
		\caption{  \label{Fig:OP} Temperature dependence of the $(0,0,4-\tau_z)$ magnetic-Bragg peak for $x=0.10$ (a), the $(0,0,4-\tau_z)$ magnetic-Bragg peak for $x=0.20$ (b), the $(0,0,2+\tau_z)$ magnetic and $(1,1,0)$ structural and magnetic-Bragg peaks for $x=0.42$ (c), and the $(0,0,2+\tau_z)$ magnetic-Bragg peak for $x=0.82$ (d).  Solid lines are guides to the eye. The lower-temperature vertical-dashed lines indicate the N\'eel temperature $T_{\text{N}}$ and the higher-temperature vertical-dashed line for $x=0.42$ indicates the Curie temperature $T_{\text{C}}$.} 
	\end{figure}
	
	Figure~\ref{Fig:OP} shows the temperature dependence of the intensity of a magnetic-Bragg peak for each sample.  These data indicate the temperature dependence of the magnetic order parameter as long as the center position or full-width-at-half maximum (FWHM)  of the peak does not change with temperature. Fits of Gaussian lineshapes to magnetic-Bragg peaks for each $x$ showed no significant shifts in the peaks' centers nor significant changes to their FWHMs over the temperature ranges measured, except for the typical larger FWHMs just below $T_{\text{N}}$ that are consistent with shorter magnetic-correlation lengths due to critical fluctuations.  The fact that the centers of the peaks did not change with temperature means that within our resolution we did not observe temperature-dependent changes of $\tau_z$. $T_{\text{N}}$ and $\tau_z$ are reported below in Table~\ref{Tab:Params}.
	
	\begin{figure}[]
		\centering
		\includegraphics[width=1.0\linewidth]{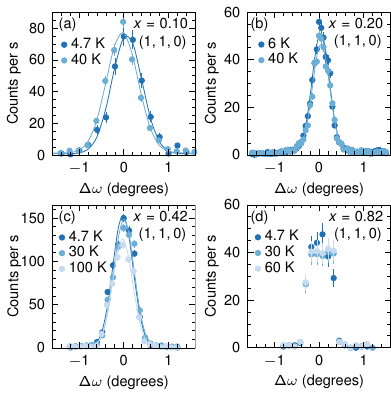}
		\caption{  \label{Fig:110rock} Data from rocking scans across the $(1,1,0)$ Bragg peak at various temperatures for (a) $x=0.10$, (b) $x=0.20$, (c) $x=0.42$, and (d) $x=0.82$. Lines are fits to Gaussian lineshapes. The shapes of the peaks for $x=0.82$ are strongly influenced by absorption. Gaussian fits for these peaks are not shown as the lineshapes are apparently similar within uncertainty for the different temperatures.} 
	\end{figure}
	
	Neutron diffraction can provide information regarding the direction of $\bm{\mu}$ since it is sensitive to the component of $\bm{\mu}$ perpendicular to the scattering vector $\mathbf{Q}$. Thus, the magnetic-Bragg peaks shown in Fig.~\ref{Fig:l_scans} for $(0,0,l)$ are consistent with $\bm{\mu}$ having a component in the $\bm{ab}$ plane.  Likewise, the previously-reported higher-temperature $c$-axis FM phase with $\bm{\mu}\parallel\mathbf{c}$ should be characterized by magnetic-Bragg peaks occurring on top of integer $(h,k,l)$ structural-Bragg peaks with nonzero values of $h$ or $k$ upon cooling below the Curie temperature $T_{\text{C}}$. 
	
	To this end, Fig.~\ref{Fig:110rock} shows base-temperature and higher-temperature rocking-scan data for the $(1,1,0)$ peak for each $x$. No significant changes in the intensity of the $(1,1,0)$ peak are seen between the high- and low-temperature data in Figs.~\ref{Fig:110rock}(a) and \ref{Fig:110rock}(d) whereas Fig.~\ref{Fig:110rock}(c) shows that the $(1,1,0)$ peak intensity increases between $T=100$ and $30$~K. This indicates the occurrence of a magnetic-Bragg peak consistent with FM order for $x=0.42$, but not for $x=0.10$ and $0.82$. Figure~\ref{Fig:l_scans}(c) shows that there is no increase in the intensity of  $(0,0,l)$ Bragg peaks for $T<T_{\text{C}}$ for $x=0.42$, which means that the FM ordering has $\bm{\mu}\parallel\mathbf{c}$.  Thus, these results all agree with the previous report of a $c$-axis FM phase for $x=0.42$ but not for $x=0.10$ and $0.82$ \cite{Sangeetha_2020}.
	
	Similar observations to those for $x=0.42$ are expected for $x=0.20$. Unfortunately, due to beam-time limitations we did not record data above the reported $T_{\text{C}}$ of $60(2)$~K \cite{Sangeetha_2020}. In addition, the rocking-scan data for the $x=0.20$ $(1,1,0)$ peak in Fig.~\ref{Fig:110rock}(b) only show a slight increase from $T=40$~K to $5$~K which is comparable to the uncertainty. Thus, from the neutron-diffraction experiments we cannot conclusively state that we observed $c$-axis FM order for $x=0.20$. On the other hand, as discussed below, single-crystal refinements made using multiple Bragg peaks indicate that $c$-axis cone order is present at $6$~K for $x=0.20$.
	
	Figure~\ref{Fig:OP}(c) shows the temperature dependence of the $(1,1,0)$ Bragg peak for $x=0.42$, which increases with decreasing temperature between $T_{\text{C}}=53(1)$~K and $T_{\text{N}}=23.6(1)$~K.  Remarkably, the intensity becomes constant below $T_{\text{N}}$. We note that the value of $T_{\text{C}}$ for $x=0.42$ determined by neutron diffraction is $5$~K lower than the value of $58(1)$~K found from magnetic susceptibility and resistance data but in line with the value of $54.3(2)$~K found by heat-capacity measurements \cite{Sangeetha_2020}. 
	
	\begin{figure}[]
		\centering
		\includegraphics[width=1.0\linewidth]{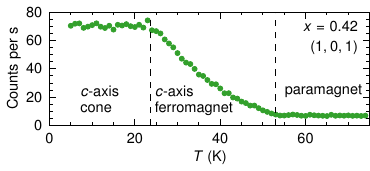}
		\caption{  \label{Fig:xp4_101}Temperature dependence of the $(1,0,1)$ structural and magnetic-Bragg peaks for $x=0.42$. } 
	\end{figure}
	
	We next examine whether the $c$-axis FM order between $T_{\text{N}}$ and $T_{\text{C}}$ for $x=0.42$ can be associated specifically with Co/Ni spins, as previously proposed \cite{Sangeetha_2020}. Looking at the reflection conditions for the $I4/mmm$ space group for the chemical lattice, we find that the special reflection condition $(h,k,l)$, $l$ even, exists for the Co/Ni crystallographic site \cite{Aroyo_2006}. This means that FM ordering of the Co/Ni spins will only cause magnetic-Bragg peaks to appear on top of structural-Bragg peaks at positions satisfying this special reflection condition.  However, our data for $x=0.42$ show that within the $c$-axis FM phase magnetic-Bragg peaks exist at positions other than those satisfying the special reflection condition. The simplest explanation is that the Eu spins exhibit $c$-axis FM order.
	
	Figure~\ref{Fig:xp4_101} demonstrates the violation of the special reflection condition for the $x=0.42$ sample by showing the temperature dependence of the intensity of the $(1,0,1)$ Bragg peak.  Similar to the data for $(1,1,0)$ shown in Fig.~\ref{Fig:OP}(c), an increase in the intensity of the $(1,0,1)$ peak develops upon cooling below $T_{\text{C}}$ consistent with the emergence of a magnetic-Bragg peak on top of a structural-Bragg peak. Also similar to the data for $(1,1,0)$ is that the intensity of the $(1,0,1)$ magnetic-Bragg peak becomes constant below $T_{\text{N}}$. This suggests that the size of the ferromagnetically-ordered moment along $\mathbf{c}$ becomes constant below $T_{\text{N}}$.
	
	To address whether the Co/Ni spins also exhibit $c$-axis FM order for $x=0.42$ and $T\leq T_{\text{C}}$, we used the integrated intensities for the $(1,1,0)$ and $(1,0,1)$ Bragg peaks recorded at $T>T_{\text{C}}$ and $T<T_{\text{N}}$ and calculated the nuclear and magnetic structure factors for each peak.  For the magnetic structure factor, we assumed $c$-axis FM order for both the Eu and Co/Ni spins. This allowed us to solve for $\mu$ for both the Eu ($\mu_{\text{Eu}}$) and Co/Ni ($\mu_{\text{Co/Ni}}$) spins. We determined that within our sensitivity $\mu_{\text{Co/Ni}}\lesssim0.2~\mu_{\text{B}}$. This result is supported by single-crystal refinements similar to those described next but which allowed for $\mu_{\text{Co/Ni}}\neq0$.
	
	\begin{figure}[]
		\centering
		\includegraphics[width=1.0\linewidth]{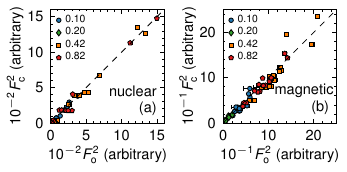}
		\caption{  \label{Fig:Refine} The square of the calculated structure factor versus the square of the observed structure factor from the single-crystal refinements for the (a) nuclear (structural) and (b) magnetic-Bragg peaks of the $x=0.10$, $0.20$, $0.42$ and $0.82$ compounds. The temperatures for the nuclear (magnetic) data are the same as those for the higher (lower) temperature data shown in Fig.~\ref{Fig:l_scans}.} 
	\end{figure}
	
	\begin{figure}[]
		\centering
		\includegraphics[width=1.0\linewidth]{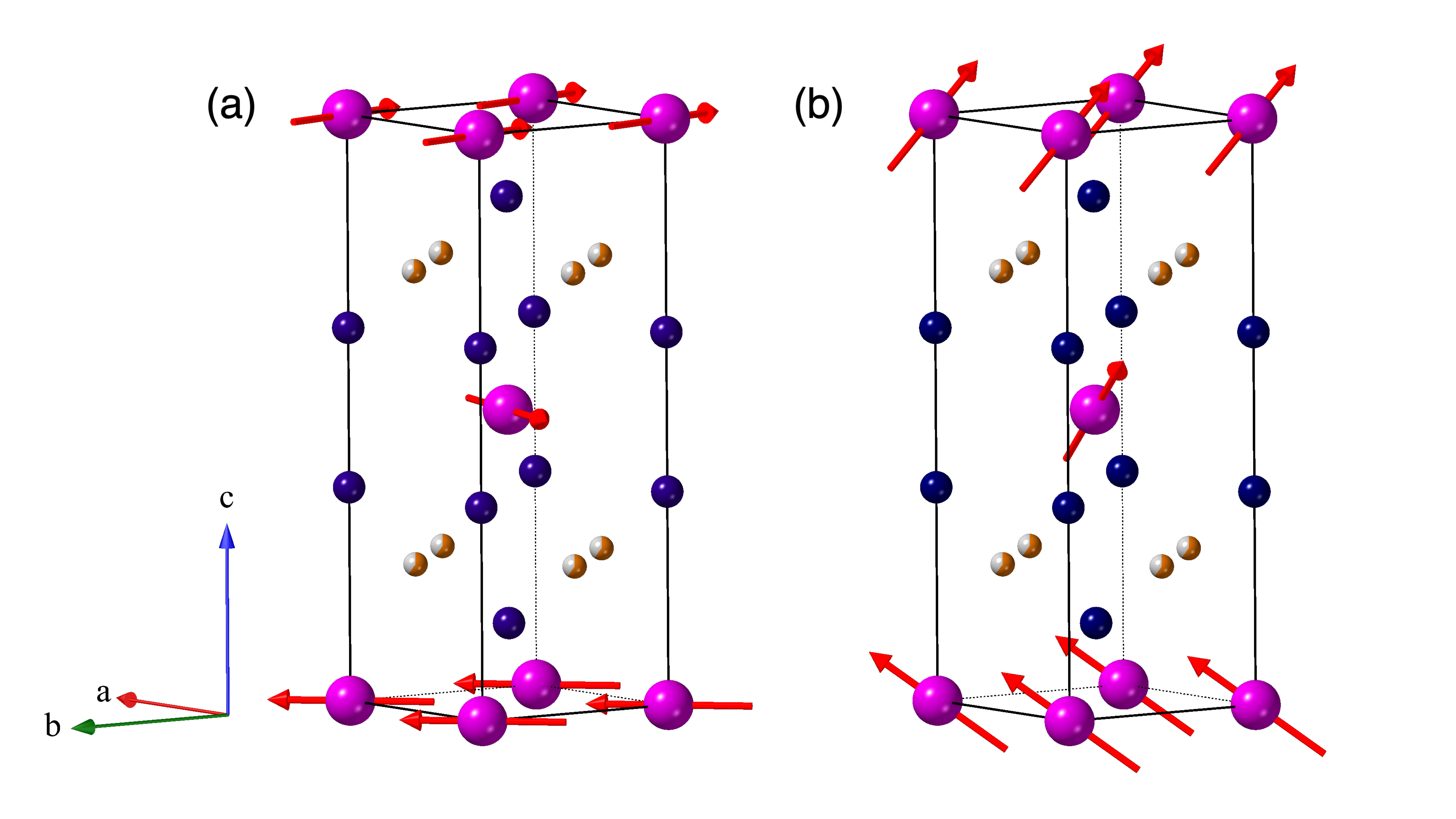}
		\caption{  \label{Fig:Mag_Struc} (a) The $c$-axis helix order of the Eu magnetic moments for $x=0.10$ and $0.82$. (b) The $c$-axis cone order of the Eu magnetic moments for $x=0.20$ and $0.42$. } 
	\end{figure}
	
	To determine the detailed structure of the AFM order for each composition measured, we used the integrated intensities of several magnetic-Bragg and structural-Bragg peaks to make single-crystal refinements.  The temperatures used for the nuclear (magnetic) data are the same as those for the higher (lower) temperature data shown in Fig.~\ref{Fig:l_scans}. We find that the data are consistent with spins within each Eu layer being ferromagnetically aligned and the orientation of each layer rotating around $\mathbf{c}$ by a turn angle of $\phi=\pi\tau_z$ between neighboring layers.  For $x=0.10$ and $0.82$, the data are consistent with a $c$-axis helix with $\bm{\mu}\perp\mathbf{c}$, similar to the AFM structures for EuCo$_{2-y}$As$_2$ \cite{Tan_2016, Ding_2017} and EuNi$_{2-y}$As$_2$ \cite{Jin_2019}.  Data for $x=0.20$ and $0.42$, on the other hand, are consistent with $c$-axis cone order. These AFM structures are shown in Fig.~\ref{Fig:Mag_Struc}. For the final refinements for $x=0.20$ and $0.42$ we assumed that $\mu_{\text{Co/Ni}}=0$, as refinements allowing for nonzero $\mu_{\text{Co/Ni}}$ yielded $\mu_{\text{Co/Ni}}=0.0(2)~\mu_{\text{B}}$, which is in agreement with our $\mu_{\text{Co/Ni}}\lesssim0.2~\mu_{\text{B}}$ result stated above.
	
	\begin{table*}
		\caption{ \label{Tab:Params}  Parameters determined from the neutron diffraction experiments. The models used for the single-crystal refinements assumed that only the Eu ions are magnetically ordered.  $T_{\text{C}}$ is the Curie temperature, $T_{\text{N}}$ is the N\'eel temperature, $\mu$ is the total ordered magnetic moment, $\mu_{\parallel c}$ is the component of $\bm{\mu}$ along $\mathbf{c}$,  $\mu_{\perp c}$ is the component of $\bm{\mu}$ in the $\mathbf{ab}$ plane, $\bm{\tau}=(0,0,\tau_z)$ is the antiferromagnetic propagation vector, the column labelled $T$ gives the temperature for the data used for the refinements, and $\chi^2$ is the usual goodness-of-fit parameter for the refinement determining the magnetic structure. As discussed in the text, the quoted uncertainties for $\mu$ are from the refinement results and likely do not reflect all of the uncertainty associated with the significant neutron absorption of the samples.} 
		\begin{ruledtabular}
			\begin{tabular}{d{1.2}cd{2.3}d{1.4}d{1.4}d{1.4}d{0.4}d{1}d{1.1}}		
				\multicolumn{1}{c}{$x$}&\multicolumn{1}{c}{$T_{\text{C}}$~(K)}&\multicolumn{1}{c}{$T_{\text{N}}$~(K)}&\multicolumn{1}{c}{$\mu$~($\mu_{\text{B}}$)}&\multicolumn{1}{c}{$\mu_{\parallel c}$~($\mu_{\text{B}}$)} &\multicolumn{1}{c}{$\mu_{\perp c}$~($\mu_{\text{B}}$)}&\multicolumn{1}{c}{$\tau_z$~(r.l.u.)}&\multicolumn{1}{c}{$T$~(K)}&\multicolumn{1}{c}{$\chi^2$}\\
				\hline
				0.0 ~\footnote[1]{Parameters are from Ref.~[\onlinecite{Tan_2016}].}& \multicolumn{1}{c}{---} & 47 & 7.26(8) & \multicolumn{1}{c}{---}&7.26(8)&0.79&5&\multicolumn{1}{c}{---} \\
				0.10&\multicolumn{1}{c}{---} & 41.7(4) & 6.5(2) & \multicolumn{1}{c}{---}&6.5(2) &0.670(1)&5&2.1\\
				0.20 & \phantom{1}60(2) \footnote[2]{Determined in Ref.~[\onlinecite{Sangeetha_2020}] from magnetic susceptibility data.}& 29.0(4) & 6.8(8)&2.9(8)& 6.2(2)&0.665(5)&6&3.9\\
				0.42 & 53(1) & 23.6(1) & 7.6(2) &3.5(2) &6.7(1)&0.703(1)&5&4.9\\ 
				0.82 &\multicolumn{1}{c}{---} & 15.0(6) & 6.9(9)& \multicolumn{1}{c}{---}&6.9(9)& 0.848(4)&5&7.3\\
				1.0~\footnote[3]{Parameters are from Ref.~[\onlinecite{Jin_2019}].}& \multicolumn{1}{c}{---} & 15.0(1) & 6.75(6) & \multicolumn{1}{c}{---}& 6.75(6)&0.92&2&\multicolumn{1}{c}{---}\\
			\end{tabular}
		\end{ruledtabular}
	\end{table*}
	
	Figure~\ref{Fig:Refine} shows the square of the calculated structure factor plotted versus the square of the observed structure factor for the nuclear [Fig.~\ref{Fig:Refine}(a)] and magnetic-Bragg peaks [Fig.~\ref{Fig:Refine}(b)] for each $x$, and parameters determined from the refinements are listed in Table~\ref{Tab:Params}. The quoted uncertainties are values from the single-crystal refinements. Overall, the refinements give satisfactory goodness-of-fit parameters, considering the significant thermal-neutron absorption of Eu. Further confidence in the results is gained by the fact that $\mu$ for each of the samples is $\approx7~\mu_{\text{B}}$, which is the value expected for Eu$^{2+}$, plus the values of $\mu_{\parallel c}$ determined for $x=0.20$ and $0.42$ are similar to the spontaneous moments found from $M(H)$ data \cite{Sangeetha_2020}.  To test the uncertainty present due to the thermal-neutron absorption and our corrections for this absorption, we made absorption corrections for varying thicknesses of the $x=0.42$ sample and repeated the single-crystal refinements.  We found that varying the value of the thickness of the sample entered into the absorption-correction algorithm by $\pm0.1$~mm can alter the determined value for $\mu$ by $\approx0.6~\mu_{\text{B}}$.
	
	\section{Discussion}
	Table~\ref{Tab:Params} reveals no clear trend in the sizes of $\mu_{\parallel c}$ and $\mu_{\perp c}$ with $x$, although the strong sample-shape dependent neutron absorption likely obfuscates small changes to the fitted values of the ordered moments.  On the other hand, it is notable that Fig.~\ref{Fig:Phase_Dia}(b) shows that the slope of $\tau_z(x)$ switches sign between $x=0.10$ and $0.20$. This is where the Tet-cT crossover is reported to occur \cite{Sangeetha_2020} and this structural change can be expected to affect $\tau_z$.
	
	For example, within a mean-field model using a $J_0$-$J_1$-$J_2$ Hamiltonian to describe $c$-axis helix ordering, $\tau_z$ determines the helix turn angle and is related to the exchange strengths through
	\begin{align}
		\tau_z=\frac{c}{2\pi d}&\cos^{-1}\left(-\frac{J_1}{4J_2}\right) \label{Eqn:tauz}\\  
		=\frac{1}{\pi}&\cos^{-1}\left(-\frac{J_1}{4J_2}\right)\ . \nonumber
	\end{align}
	Here, $J_1$ and $J_2$ are the effective interlayer-exchange strengths between nearest-neighboring and next-nearest-neighboring Eu layers, $d=c/2$ is the spacing between Eu layers, and $J_0$ is the effective FM intralayer exchange that ferromagnetically aligns spins in each Eu layer \cite{Johnston_2012,Johnston_2019}. Using this model to describe the situation below $T_{\text{N}}$ and assuming that only the Eu spins are ordered, the apparent minimum for $\tau_z(x)$ seen in Fig.~\ref{Fig:Phase_Dia}(b) would be due to changes of the interlayer-exchange strengths associated with enhanced interlayer As-As bonding as the compound enters the cT phase. The subsequent increase in $\tau_z(x)$ with increasing $x$ would then presumably be due to further changes in the interlayer-exchange strengths associated with the As-As bonding; the As atoms sit at the $4e$ site of the chemical unit cell with coordinates $(0,0,\pm z_{\text{As}})$, and $z_{\text{As}}$ monotonically increases with $x$ past the Tet-cT crossover \cite{Sangeetha_2020}. Possible explanations for the emergence of a component of $\bm{\mu}$ parallel to $\mathbf{c}$ for $0.20\lesssim x<0.75$ and the resulting $c$-axis cone order, however, require additional considerations.
	
	For a magnetic rare earth with $L\neq0$, the CEF splitting of a rare-earth cation's ground-state angular-momentum multiplet would be expected to establish some magnetic anisotropy.  Since the CEF splitting depends on the details of the anions surrounding the rare earth, it would then be interesting to consider if the Tet-cT crossover could sufficiently affect the CEF splitting to change the ground-state magnetic anisotropy. However, for Eu(Co$_{1-x}$Ni$_x$)$_{2-y}$As$_2$  the Eu$^{2+}$ cation has $L=0$. This means that no CEF splitting of the ground state $J=7/2$ multiplet will occur.  Thus, there must be other mechanisms associated with the $4f$ orbitals and spins of Eu(Co$_{1-x}$Ni$_x$)$_{2-y}$As$_{2}$, such as anisotropic exchange or strong magnetic-dipole interactions, causing the emergence of the $\mathbf{c}$ component to $\bm{\mu}_{\text{Eu}}$ upon entry into the cT phase, as well as the disappearance of the $\mathbf{c}$ component of $\bm{\mu}_{\text{Eu}}$ for $x\geq0.75$.
	
	Another possible explanation is that interactions between $c$-axis FM ordered Co/Ni spins and the Eu spins cause the emergence of a $c$-axis ordered Eu moment below $T_{\text{C}}$ for those values of $x$ with a $c$-axis FM phase. To consider this explanation, we first examine whether evidence exists for itinerant-FM ordering of the Co/Ni spins in Eu(Co$_{1-x}$Ni$_x$)$_{2-y}$As$_{2}$. First, as noted above, our neutron diffraction data put a limit of $\mu_{\text{Co/Ni}}\lesssim0.2~\mu_{\text{B}}$ for the Co/Ni ordered moment within the $c$-axis FM phase, and such a weak (but finite) moment would be generally consistent with itinerant-FM order.
	
	Next, some previously reported results also support the possible occurrence of itinerant-FM ordering of the Co/Ni spins in Eu(Co$_{1-x}$Ni$_x$)$_{2-y}$As$_2$. These results are: ($1$) features in the magnetic susceptibility, M\"ossbauer spectroscopy, and heat capacity data for Eu(Co$_{1-x}$Ni$_x$)$_{2-y}$As$_2$ \cite{Sangeetha_2020}; ($2$) pressure-induced itinerant-FM order found for EuCo$_{2-y}$As$_2$ \cite{Tan_2016} and structurally-related EuCo$_{2}$P$_2$ \cite{Chefki_1998} which appears to arise from an increase in the Co $3d$ density of states in the cT phase.  Regarding point ($1$), the nonmonotonic change to the effective magnetic moment with $x$, step-like features at $T_{\text{C}}$ in temperature-dependent heat capacity data for $0.20\le x<0.75$, and M\"ossbauer data consistent with a molecular field due to an effective exchange interaction between ordered Co/Ni and Eu spins for $x=0.20$ and $0.65$ are all reported in Ref.~[\onlinecite{Sangeetha_2020}] as being consistent with itinerant-FM order of the Co/Ni spins within the $c$-axis FM phase.
	
	Putting point ($2$) in the context of Eu(Co$_{1-x}$Ni$_x$)$_{2-y}$As$_2$ takes some elucidation.  As mentioned in the Introduction, x-ray spectroscopy results for EuCo$_{2-y}$As$_2$ show that applied pressure causes a continuous Tet-cT transition at a critical pressure of $p_{\text{c}}\approx4.6$~GPa which is accompanied by itinerant-FM ordering of the Co spins \cite{Tan_2016}. These experiments also found that the valence of Eu increases across the transition, reaching an average value of $2.25$ by $12.6$~GPa.  This valence increase for the cT phase is expected from density functional theory (DFT) calculations to increase the density of states at the Fermi energy $\rho(E_{\text{F}})$ by increasing the $3d$ density of states associated with Co \cite{Tan_2016}. This increase in $\rho(E_{\text{F}})$ for the cT phase can explain the emergence of the itinerant-FM order via the Stoner theory, which states that itinerant-FM order occurs when the Stoner parameter $\alpha_0=\rho(E_{\text{F}}) I$ is $\alpha_0>1$. Here, $I$ is the normalized average effective Coulomb repulsion between electrons on the same site \cite{Moriya_1985, Takahashi_2013}. Thus, the increase in $\rho(E_{\text{F}})$ for the cT phase calculated by DFT can explain the presence of itinerant-FM ordering of the Co spins above $p_{\text{c}}$ for EuCo$_{2-y}$As$_2$.
	
	The same mechanism has been invoked for the emergence of magnetic ordering of the Co spins in the pressure-induced cT phase of EuCo$_{2}$P$_2$ \cite{Chefki_1998}.  However, for EuCo$_{2}$P$_2$, the high-pressure magnetic order is AFM overall, reported to consist of Co layers with ferromagnetically aligned spins and the Co layers stacked along $\mathbf{c}$ in a $+$$+$$+$$-$$-$$-$ sequence with $+$ and $-$ indicating opposite directions of the ordered magnetic moment of a layer \cite{Chefki_1998}. This order is not inconsistent with the presence of itinerant ferromagnetism and also points to the existence of competing FM and AFM interlayer interactions.  This competition may be more readily revealed in EuCo$_{2}$P$_2$ than EuCo$_{2-y}$As$_2$ because $\bm{\mu}_{\text{Eu}}$ is quenched in the cT phase for EuCo$_{2}$P$_2$ \cite{Chefki_1998} whereas it remains finite for EuCo$_{2-y}$As$_2$ \cite{Tan_2016}. 
	
	Similar to the above discussion of itinerant-FM order in the cT phases of EuCo$_{2-y}$As$_2$ and EuCo$_{2}$P$_2$, we can examine if $\rho(E_{\text{F}})$ increases at the Tet-Ct crossover in Eu(Co$_{1-x}$Ni$_x$)$_{2-y}$As$_2$ and consider the correlation between the existence of $c$-axis FM ordering and changes to $\rho(E_{\text{F}})$ with $x$. This is done by using the previously reported values for the Sommerfeld coefficient $\gamma=\pi^2k_{\text{B}}^2\rho(E_{\text{F}})/3$. From the results in Ref.~[\onlinecite{Sangeetha_2020}], we see that an increase in $\rho(E_{\text{F}})$ from $6.3(8)$ to $13.9(4)~\text{states}/\text{eV-f.u.}$\ (where f.u.\ stands for formula unit) occurs between $x=0$ and $0.20$.  $\rho(E_{\text{F}})$ then decreases with further increasing $x$, becoming $\rho(E_{\text{F}}) = 9.0(3)~\text{states}/\text{eV-f.u.}$\ for $x=0.82$. From Ref.~[\onlinecite{Sangeetha_2019}], we find a still smaller value of $\rho(E_{\text{F}}) \sim 2~\text{states}/\text{eV-f.u.}$ for $x=1$.  Thus, within the Stoner picture, the increase in $\rho(E_{\text{F}})$ between $x=0$ and $0.20$ and the observation of $c$-axis FM order for $0.20\lesssim x<0.75$ could be consistent with the onset of itinerant-FM order accompanying the Tet-cT crossover due to the Stoner condition ($\alpha_0>1$) becoming satisfied. The decrease in $\rho(E_{\text{F}})$ with further increasing $x$ would be consistent with the eventual disappearance of such order. Missing from this analysis, however, is that $I$ can change as Co is replaced by Ni, which would also affect $\alpha_0$.
	
	Assuming that itinerant-FM ordering of the Co/Ni spins exists for the $c$-axis FM phase, we hypothesize that the molecular field due to such order causes the emergence of a component of $\bm{\mu}_{\text{Eu}}\parallel\mathbf{c}$ for $0.20\lesssim x < 0.75$ below $T_{\text{C}}$. Indeed, as noted in the Results, the intensities of the $(1,1,0)$ and $(1,0,1)$ magnetic-Bragg peaks shown in Figs.~\ref{Fig:OP}(c) and \ref{Fig:xp4_101} only increase upon cooling from $T_{\text{C}}$ to $T_{\text{N}}$. This means that the size of $\mu_{\text{Co/Ni}}$ and $\bm{\mu}_{\text{Eu}}\parallel\mathbf{c}$ only grow upon cooling from $T_{\text{C}}$ to $T_{\text{N}}$ and remain constant below $T_{\text{N}}$. Next, based on the fact that without a $c$-axis FM phase the Eu sublattice energetically favors $c$-axis helix ordering, we propose that for $0.20\lesssim x < 0.75$ the $ab$-plane anisotropy of the Eu spins is partially overcome by the molecular field due to the $c$-axis FM ordering of the Co/Ni spins, resulting in the $c$-axis cone phase below $T_{\text{N}}$. Element-specific magnetic characterization of the electrons associated with the Co/Ni and Eu spins by specialized measurements such as x-ray magnetic circular dichroism and x-ray resonant magnetic scattering on Eu(Co$_{1-x}$Ni$_x$)$_{2-y}$As$_2$ to examine any FM order present and changes in valence may prove enlightening. 
	
	Finally, we note that the neutron-diffraction experiments did not observe the multiple transitions around $T_{\text{N}}$ observed by M\"ossbauer experiments.  These transitions were interpreted as the AFM ordering of the Eu magnetic moments being a spin-density wave (SDW) which squares up with cooling \cite{Sangeetha_2020}.  Our diffraction experiments cannot differentiate between helical or SDW ordering due to the presumed presence of magnetic domains. However, within our resolution and sensitivity, we did not observe a change in $\bm{\tau}$ with temperature which would be suggestive of a SDW nor did we observe magnetic-Bragg peaks corresponding to odd harmonics of $\tau$ (i.e.\ $3\tau$) which would correspond to squaring up of a SDW. X-ray resonant magnetic scattering would likely be a more definitive probe to address the M\"ossbauer results due to the technique's typically higher momentum-transfer resolution than neutron diffraction and the fact that the thermal-neutron absorption of Eu limits the sensitivity of neutron diffraction to changes in $\mu$.
	
	\section{Conclusion}
	We have reported results from neutron-diffraction experiments on single-crystal samples of Eu(Co$_{1-x}$Ni$_x$)$_{2-y}$As$_2$ that have directly determined the magnetic ordering of the Eu spins across the previously reported magnetic phase diagram \cite{Sangeetha_2020}. For $x=0.10$ and $0.82$, $c$-axis helix ordering of the Eu spins exists below $T_{\text{N}}=41.7(4)$~K and $15.0(6)$~K, respectively.  For $x=0.20$ and $0.42$, higher-temperature $c$-axis FM ordering of the Eu spins occurs  below $T_{\text{C}}=60(2)$~K \cite{Sangeetha_2020} and $53(1)$~K, respectively, and  $c$-axis cone order of the Eu spins sets in below $T_{\text{N}}=29.0(4)$~K and $23.6(1)$~K, respectively. Our single-crystal refinements using the diffraction data found that $\mu\approx7~\mu_{\text{B}}$ at $T\approx5$~K for all of the compositions measured and we found that $\tau_z(x)$ first decreases then increases with increasing $x$, with the minimum in $\tau_z(x)$ accompanying the continuous Tet-cT structural-phase transition.
	
	The discovery of $c$-axis FM ordering of the Eu spins for $x=0.20$ and $0.42$ is a result of our neutron diffraction experiments' ability to examine the existence of magnetic-Bragg peaks in regards to the extinction conditions for the magnetic structure factor. We were unable, however, to directly confirm whether the Co/Ni spins exhibit the previously reported itinerant-FM order \cite{Sangeetha_2020} but put an upper limit of $\mu_{\text{Co/Ni}}\lesssim0.2~\mu_{\text{B}}$ for the ordered moment associated with such order. Contrary to the previously reported phase diagram \cite{Sangeetha_2020}, we also found no evidence for a second AFM propagation vector for $x=0.10$.
	
	In analyzing our results, we discussed how $\tau_z(x)$ may be affected by the Tet-cT crossover using a mean-field model. We also discussed how $\rho(E_{\text{F}})$ determined from the Sommerfeld coefficient appears to reach a maximum at the Tet-cT crossover and may be responsible for the onset of Stoner-type itinerant ferromagnetism of the Co/Ni spins. We postulated that the occurrence of $c$-axis FM order of the Co/Ni spins could provide a molecular field that results in a $c$-axis ordered component for the Eu spins, explaining the higher-temperature Eu $c$-axis FM and $c$-axis cone ordering for intermediate $x$. Element specific magnetic measurements such as x-ray magnetic circular dichroism and x-ray resonant magnetic scattering should prove fruitful.
	
	\begin{acknowledgments}
		We are grateful for conversations with Huibo~Cao, Andreas~Kreyssig and David~Vaknin. Work at the Ames National Laboratory was supported by the U.~S.~Department of Energy (USDOE), Basic Energy Sciences, Division of Materials Sciences \& Engineering, under Contract No.~DE-AC$02$-$07$CH$11358$. 
	\end{acknowledgments}

	\bibliography{Ni_doped_EuCo2As2.bib}
	
\end{document}